\documentstyle[preprint,prc,aps,eqsecnum,epsf]{revtex}
\tighten
\begin{document}
\draft
\title{Theoretical Study of the $^3$He($\mu^-$,$\nu_\mu$)$^3$H Capture} 
\author{
L.E. Marcucci$^{(a,b)}$, R. Schiavilla$^{(c,d)}$, S. Rosati$^{(a,b)}$ , 
A. Kievsky$^{(b,a)}$, and M. Viviani$^{(b,a)}$ }
\address{
(a) Department of Physics, University of Pisa, I-56100 Pisa, Italy\\
(b) INFN, Sezione di Pisa, I-56100 Pisa, Italy\\
(c) Department of Physics, Old Dominion University, Norfolk, Virginia 23529, USA \\
(d) Jefferson Lab, Newport News, Virginia 23606, USA 
}
\date{\today}
\maketitle

\begin{abstract}
The $^3$He($\mu^-$,$\nu_\mu$)$^3$H weak capture is studied
using correlated-hyperspherical-harmonics wave functions,
obtained from realistic Hamiltonians consisting of the Argonne
$v_{14}$ or Argonne $v_{18}$ two-nucleon, and Tucson-Melbourne
or Urbana-IX three-nucleon interactions.  The nuclear weak
charge and current operators have vector and axial-vector
components with one- and two-body contributions.  The axial-vector 
current includes the nucleon and $\Delta$ induced pseudo-scalar 
terms, with coupling constants $g_{PS}$ and $g_{PS}^*$ derived
from pion-pole dominance and PCAC.  The strength of the leading two-body
operator is adjusted to reproduce the Gamow-Teller matrix element
in tritium $\beta$-decay.  The calculated total capture rate is
within $\sim$0.5 \% of the most recent measurement, $1496\pm 4$ sec$^{-1}$.
The predictions for the capture rate and angular correlation parameters
$A_v$, $A_t$, and $A_\Delta$ are found to be only very weakly dependent
on the model input Hamiltonian.  The variation of the observables with
$g_{PS}$ and $g_{PS}^*$ and the theoretical uncertainties deriving
from the model-dependent procedure used to constrain the axial current
are investigated. 
\end{abstract}

\section{Introduction}
\label{sec:intro}
The $\mu^-$ weak capture on $^3$He can occur through three 
different hadronic channels:
\begin{eqnarray}
\mu^- + ^3\!{\rm He} &\rightarrow& ^3{\rm H} +\nu_\mu ~~~(70\%)\ , 
\label{eq:mucap} \\
\mu^- + ^3\!{\rm He} &\rightarrow& n\,+\,d\,+\nu_\mu ~~~(20\%)\ , 
\label{eq:mcnd} \\
\mu^- + ^3\!{\rm He} &\rightarrow& n\,+\,n\,+p\,+\nu_\mu ~~~(10\%)\ . 
\label{eq:mcnnp} 
\end{eqnarray}
The focus of the present work is on the first process.  Some of 
the nuclear physics issues in muon capture have been reviewed
recently in Ref.~\cite{Mea01}.

The reaction~(\ref{eq:mucap}) has been extensively
studied through the years, both experimentally and
theoretically.  Measurements of the total capture
rate have been performed since the early sixties~\cite{Fol63,Aue65,Cla65}
up to until recently.  The latest very precise
experimental determination of this observable~\cite{Ack98},
$1496\pm 4$ sec$^{-1}$, is consistent with the
earlier measurements, the latter having
considerably larger uncertainties, however.

Theoretical studies of reaction~(\ref{eq:mucap}) 
have been carried out within two different frameworks: 
the so-called \lq\lq elementary particle method\rq\rq
(EPM) and the fully microscopic approach. The EPM,
first developed by Kim and Primakoff~\cite{Kim65}, 
is essentially a phenomenological approach, which 
parameterizes the nuclear (charge-changing) weak
current in terms of the trinucleon form factors, 
in analogy to the
nucleon weak current, and then attempts to 
derive these from other experiments.  Within the
EPM, it was shown in Ref.~\cite{Hwa78} 
that, if the hyperfine structure of the
$\mu^-\, ^3$He system is taken into account
and the direction of the recoiling triton can
be detected, there are, in addition to the
capture rate, other observables, i.e. angular
correlation parameters, which are more sensitive
than the capture rate itself to the value of the
nucleon pseudo-scalar axial coupling constant $g_{PS}$.
Indeed, the possibility of determining $g_{PS}$
from measurements of muon capture observables is
one of the motivations for the interest 
that this process has generated over the years.
Recently, one attempt has been made to
measure the angular correlation
parameter $A_v$~\cite{Sou98}, though the experimental
result, which to our knowledge represents
the first significant measurement of this
observable, is affected by large
systematic uncertainties.  Therefore, a
comparison between theory and experiment
will not be particularly meaningful for $A_v$.
Experimental results with an improved accuracy
are highly desirable.

In contrast, the fully microscopic approach is
based on: i) $^3$H and $^3$He wave functions as
accurate as possible, to reduce uncertainties
related to nuclear structure; ii) a realistic
model for the nuclear weak current and charge
operators.  The first microscopic calculation
of reaction~(\ref{eq:mucap}) was performed by
Peterson in 1968~\cite{Pet68}, and was reconsidered
and improved by Phillips and collaborators in
1974~\cite{Phi74}.  These studies, however, used
nuclear wave functions which were approximate,
and retained in the nuclear weak transition operators
only single-nucleon terms, the impulse approximation (IA). 

In the early nineties, the muon capture on $^3$He,
including the total rate and angular correlation
parameters mentioned above, have been extensively
investigated, within the fully microscopic framework,
by Congleton and Fearing~\cite{Con92}, and
Congleton and Truhl\`{\i}k~\cite{Con95}.  The most
significant improvements in these studies, relative
to those of the late sixties~\cite{Pet68} and early
seventies~\cite{Phi74}, are in the more accurate treatment
of the trinucleon wave functions and of the 
weak interaction. The nuclear wave functions have 
been obtained from a realistic Hamiltonian based 
on the Argonne $v_{14}$ two-nucleon~\cite{Wir84}
and Tucson-Melbourne three-nucleon~\cite{Coo79} interactions,
using the rearrangement coupled-channel method~\cite{Kam89}.

The study in Ref.~\cite{Con92} used the IA form
of the nuclear weak current, and emphasized 
the need to go beyond single-nucleon 
contributions for a realistic description of the process.
This next step was carried out in Ref.~\cite{Con95}, where 
a model for two-body components in the nuclear weak
current was explicitly constructed.  The calculated
capture rate is in good agreement with the measured
value, although the theoretical prediction suffers
from a 2 \% uncertainty, which is rather large compared to
the experimental error, and mostly arises from poor
knowledge of some of the coupling constants and
cutoff parameters entering the axial current. 

The present work sharpens and updates that
of Ref.~\cite{Con95}.  Improvements in the
modeling of two- and three-nucleon interactions
and the nuclear weak current make the re-examination
of process~(\ref{eq:mucap}) especially timely.  The
initial and final state wave functions have been obtained, using the 
correlated-hyperspherical-harmonics method, from a nuclear 
Hamiltonian which consists
of the Argonne $v_{18}$ two-nucleon~\cite{Wir95} and
Urbana-IX three-nucleon~\cite{Pud95} interactions.
To make contact with the study of Ref.~\cite{Con95},
however, and to have some estimate of the model
dependence of the results, the older Argonne $v_{14}$
two-nucleon and Tucson-Melbourne
three-nucleon interaction models have
also been used.  Both these Hamiltonians reproduce
the experimental binding energies and charge radii
of the trinucleon systems.

The model for the nuclear weak current used in the present work 
has been developed in Refs.~\cite{Sch98,Mar00,Mar01}.  However,
two additional contributions have been included: the one-body
term associated with the induced pseudo-scalar charge operator
of the nucleon, and the induced pseudo-scalar two-body term in
the $N\Delta$-transition axial current.  Both contributions are
of order $O(q^2/m^2)$, where $q$ is the momentum transfer in the
process and $m$ is the nucleon mass.  They were neglected
in the proton weak capture reactions studied in Refs.~\cite{Sch98,Mar00,Mar01},
for which $q \ll m$.
A brief description of these operators is given in Sec.~\ref{sec:wcc}. 

Some of the differences between the model for the nuclear weak
current of Ref.~\cite{Con95} and that adopted here should be noted.
It is well known by now that the axial current associated with
$\Delta$ excitation is the dominant (axial) two-body mechanism.
In the present work, its strength, i.e. the $N$$\Delta$-transition
axial coupling constant $g_A^*$, has been determined by fitting the
measured Gamow-Teller matrix element in tritium $\beta$-decay.  
The inherent model dependence of this procedure has been shown to be
very weak in studies of the proton weak captures on $^1$H~\cite{Sch98}
and $^3$He~\cite{Mar01}.  In Ref.~\cite{Sch98} predictions
for the $^1$H($p$,$e^+ \nu_e$)$^2$H cross section, obtained
with a variety of modern high-quality two-nucleon interactions,
differed by significantly less than  1 \%, once the coupling constant
$g_A^*$ had been fixed as described above within each given model
Hamiltonian (for further discussion of this point as well as
of the reasons for such a weak model dependence, see
Ref.~\cite{Sch98}).  In Ref.~\cite{Con95}, on the other hand,
$g_A^*$ is related to the $\pi$$N$$\Delta$ coupling constant
$f_{\pi N \Delta}$, and values ranging from the quark-model
to the Skyrme-soliton model predictions are used for $f_{\pi N \Delta}$.
 
There are additional differences in the detailed
form of the pion range operators, which in Ref.~\cite{Con95}
were derived from a phenomenological chiral Lagrangian
containing contributions from $\pi$- and $A_1$-pole mediated
currents.  Congleton and Truhl\`{\i}k, though, ignored
$\rho$-meson contributions both in the axial-vector and
vector sectors of the weak current.  These are retained in
the present work.  However, as it is clear from Ref.~\cite{Con95}
and also from Sec.~\ref{sec:res} below, these differences have
little numerical impact on the calculated muon capture observables.

Finally, the induced pseudo-scalar term in the $\Delta$
axial current is ignored in Ref.~\cite{Con95}, while
here it is determined using pion-pole dominance and
the partially-conserved-axial-current (PCAC) hypothesis.
The induced pseudo-scalar coupling constant $g_{PS}^*$ is
related to $g_A^*$ via the (extended) Goldberger-Treiman
relation~\cite{Hem95}.

A crucial issue, though, remains to be addressed,
namely the extent to which the present model for
the nuclear weak current is successful in predicting
observed weak transitions (note that
the cross sections of the proton weak capture
processes mentioned above are not known experimentally).
The present work fulfils this need by showing
that the calculated rate for $\mu^-$ capture on $^3$He
is in excellent agreement with the measured value.

This manuscript falls into five sections.  In
Sec.~\ref{sec:thfm} explicit expressions for the
rate and angular correlation parameters are derived
in terms of reduced matrix elements of multipole
operators, while in Sec.~\ref{sec:wcc} the model
for the weak current is succintly described.
The results are presented and discussed in Sec.~\ref{sec:res},
and some concluding remarks are given in Sec.~\ref{sec:setc}.  
\section{Observables}
\label{sec:thfm}

The muon capture on $^3$He is induced by the
weak interaction Hamiltonian~\cite{Wal75,Wal95}

\begin{equation}
H_{W}={G_{V}\over{\sqrt{2}}} \int {\rm d}{\bf x}
\, l_{\sigma}({\bf x}) j^{\sigma}({\bf x}) \ , \label{eq:hw}
\end{equation}
where $G_{V}$ is the Fermi coupling constant,
$G_{V}$=1.14939 $\times 10^{-5}$ GeV$^{-2}$~\cite{Har90},
and $l_\sigma$ and $j^\sigma$ are the leptonic and
hadronic current densities, respectively.  The former
is given by

\begin{equation}
l_{\sigma}({\bf x}) =\, 
{\rm e}^{-{\rm i} {\bf k}_\nu \cdot {\bf x} } \,
{\overline{u}}({\bf k}_\nu,h_\nu)\,\gamma_{\sigma}\, (1-\gamma_5)
\psi_{\mu}({\bf x},s_{\mu}) \>\>\>,
\label{eq:lepc}
\end{equation}
where $\psi_\mu({\bf x},s_\mu)$ is the ground-state
wave function of the muon in the Coulomb field of the $^3$He
nucleus, and $u({\bf k}_\nu,h_\nu)$ is the spinor of
a muon neutrino with momentum ${\bf k}_\nu$, energy
$E_\nu$ (=$k_\nu$), and helicity $h_\nu$.  While in principle
the relativistic solution of the Dirac equation could be
used, in practice it suffices to approximate

\begin{equation}
\psi_\mu({\bf x},s_\mu) \simeq 
\psi_{1s}(x) \chi(s_\mu) \equiv 
\psi_{1s}(x) u({\bf k}_\mu,s_\mu) 
\qquad {\bf k}_\mu \rightarrow 0 \>\>\>,
\end{equation}
since the muon velocity
$v_\mu \simeq Z \alpha \ll 1$ ($\alpha$ is
the fine-structure constant and $Z$=2).  Here
$\psi_{1s}(x)$ is the $1s$ solution of the Schr\"odinger
equation and, since the muon is essentially at rest,
it is justified to replace the two-component
spin state $\chi(s_\mu)$ with the four-component
spinor $u({\bf k}_\mu,s_\mu)$
in the limit ${\bf k}_\mu \rightarrow 0$.  This will
allow us to use standard techniques to carry out
the spin sum over $s_\mu$ at a later stage. 

In order to account for the hyperfine structure
in the initial system, the muon and $^3$He
spins are coupled to states with total spin $f$ equal to
0 or 1.  The transition amplitude can then be conveniently
written as

\begin{eqnarray}
T_W (f,f_z;s^\prime_{3},h_\nu) &\equiv&
\langle ^3{\rm H},s^\prime_{3}; \nu, h_\nu \,|\, H_W \,|\,
(\mu,^3\!{\rm He});f,f_z \rangle
\nonumber \\
&\simeq &{G_V \over \sqrt{2}} \psi_{1s}^{\rm av}
\sum_{s_\mu,s_3}
\langle {1 \over 2}s_{\mu}, {1 \over 2} s_3 | f,f_z \rangle\,
l_\sigma(h_\nu,\,s_\mu)\,
\langle ^3{\rm H},s^\prime_{3} | j^{\sigma}({\bf q}) |
^3{\rm He}, s_3\rangle \ , \label{eq:hffz}
\end{eqnarray}
where 

\begin{equation}
l_\sigma(h_\nu,\,s_\mu) \equiv
{\overline{u}}({\bf k}_\nu,h_\nu)\,\gamma_{\sigma}\, (1-\gamma_5)
u({\bf k}_\mu,s_{\mu}) \>\>\>,
\end{equation}
and the Fourier transform of the nuclear weak current
has been introduced as

\begin{equation}
j^\sigma({\bf q})=\int {\rm d}{\bf x}\,
{\rm e}^{ {\rm i}{\bf q} \cdot {\bf x} }\,j^\sigma({\bf x})
\equiv (\rho({\bf q}),{\bf j}({\bf q}))
\label{eq:jvq} \>\>\>,
\end{equation}
with the leptonic momentum transfer ${\bf q}$ defined
as ${\bf q} = {\bf k}_\mu-{\bf k}_\nu \simeq -{\bf k}_\nu$.
The Bohr radius of the muonic atom in the ground state
is about 130 fm, i.e.\ much larger than the nuclear radius, and
it is therefore well justified to factor out $\psi_{1s}(x)$
from the matrix element of $j^{\sigma}({\bf q})$
between the trinucleon ground states, by approximating it
as~\cite{Wal75,Wal95}

\begin{equation}
|\psi_{1s}^{\rm av}|^2 \equiv\,  {\cal {R}} |\psi_{1s}(0)|^2\,=\,
{\cal {R}}\,{(2\,\alpha\, m_{r})^3\over \pi} \ ,
\label{eq:psimu}
\end{equation}
where $\psi_{1s}(0)$ denotes the Bohr wave function
evaluated at the origin for a point charge $2 e$, 
$m_r$ is the reduced mass of the $\mu^-\,^3$He system, 
and the factor ${\cal {R}}$ approximately accounts
for the finite extent of the nuclear charge
distribution~\cite{Wal75,Wal95}. The value
${\cal R}$=0.98 is used here~\cite{Con92}. 
 
Standard techniques~\cite{Mar01,Wal95} are
now used to carry out the multipole expansion
of the weak charge ($\rho({\bf q})$) and current
(${\bf j}({\bf q})$) operators in the general
case in which $\theta$ is the angle between the
spin quantization axis (the $\hat{\bf z}$-axis)
and the leptonic momentum transfer ${\bf q}$:

\begin{equation}
\langle ^3{\rm H},s^\prime_{3} | \rho({\bf q}) |
^3{\rm He}, s_3 \rangle = \sqrt{2\pi}
\sum_{l=0,1}\sqrt{2l + 1}\,{\rm i}^l \,
d_{m,0}^l(-\theta)\,
\langle {1 \over 2}s_{3}, l\,m | {1 \over 2}s^\prime_{3}\rangle
\, C_{l}(q) \ , \label{eq:c} 
\end{equation}
\begin{equation}
\langle ^3{\rm H},s^\prime_{3} | j_z({\bf q}) |
^3{\rm He}, s_3 \rangle =-\sqrt{2\pi}
\sum_{l=0,1}\sqrt{2l + 1}\,{\rm i}^l \,
d_{m,0}^l(-\theta)\,
\langle {1 \over 2}s_{3}, l\,m | {1 \over 2}s^\prime_{3}\rangle
\, L_{l}(q) \ , \label{eq:l} 
\end{equation}
\begin{equation}
\langle ^3{\rm H}, s^\prime_{3} | j_\lambda({\bf q}) |
^3{\rm He}, s_3 \rangle = \sqrt{3 \pi}
\, \,{\rm i}  \, d_{m,-\lambda}^1(-\theta)\,
\langle {1 \over 2}s_{3}, l\,m | {1 \over 2}s^\prime_{3}\rangle
[-\lambda M_1(q)+E_1(q)] \ , \label{eq:em}
\end{equation}
where $m$=$s_3^\prime-s_3$,
$\lambda$=$\pm 1$, and $C_l$, $L_l$, $E_l$ and $M_l$
denote the reduced matrix elements (RME's) of the Coulomb $(C)$, longitudinal
$(L)$, transverse electric $(E)$ and transverse magnetic $(M)$ multipole
operators, as defined in Refs.~\cite{Mar01,Wal75,Wal95}.  
The $d^l_{m,m^\prime}$
are rotation matrices in the standard notation of Ref.~\cite{Edm57}.
Since the weak charge and current operators have
scalar/polar-vector $(V)$ and pseudo-scalar/axial-vector $(A)$
components, each multipole consists of the sum of $V$ and $A$ terms,
having opposite parity under space inversions~\cite{Mar01}.
Parity and angular-momentum selection rules restrict
the contributing RME's to $C_0(V)$,
$C_1(A)$, $L_0(V)$, $L_1(A)$, $E_1(A)$ and $M_1(V)$
in the $^3$He($\mu^-$, $\nu_\mu$)$^3$H process.

When the triton polarization is not detected,
the differential capture rate for the reaction~(\ref{eq:mucap}) is given by

\begin{equation}
{\rm d}\Gamma=2\pi\,\, \delta\Big(m_{\mu}+m_\tau-E_\nu - 
\sqrt{m_t^2+k_\nu^2}\,\Big)
\,|\overline{T_W}|^2 {{\rm d}{\bf k}_\nu \over {(2\pi)^3}} \ ,
\label{eq:dg}
\end{equation}
where $m_\mu$, $m_\tau$, and $m_t$ are the rest masses
of the muon, $^3$He, and $^3$H, respectively, and the
binding energy of the muonic atom has been neglected.
Note that the following definition has been introduced:

\begin{equation}
|\overline{T_W}|^2 = \sum_{s_3^\prime , h_\nu}\sum_{f , f_z}
P(f,\,f_z) |T_W(f,f_z;s^\prime_3, h_\nu)|^2 \ ,
\label{eq:hw2}
\end{equation}
where $P(f,\,f_z)$ is the probability of
finding the $\mu^-\,^3$He system
in the total-spin state $|f\,f_z\rangle$.
Integrating over the neutrino energy, the differential
capture rate reduces to:

\begin{equation}
{{\rm d}\Gamma \over {\rm d}(\cos\theta)} = {1 \over 2}\Gamma_0
\Bigg[ 1+A_v P_v \cos\theta + A_t P_t \left({3 \over 2}\cos^2\theta 
-{1 \over 2}\right) + A_\Delta P_\Delta  \Bigg] \ , \label{eq:dgd}
\end{equation}
where the total capture rate $\Gamma_0$ reads 

\begin{equation}
\Gamma_0 = G_V^2\, E_\nu^2\, \left(1-{E_\nu \over m_t}\right)
\,|\psi_{1s}^{\rm av}|^2\, \overline{\Gamma}_0 \ , \label{eq:g0}
\end{equation}
with $\overline{\Gamma}_0$ denoting
the following combination of RME's

\begin{equation}
\overline{\Gamma}_0\equiv |C_0(V)-L_0(V)|^2\,+\,|C_1(A)-L_1(A)|^2
\,+\,|M_1(V)-E_1(A)|^2 \ .
\label{eq:gtilde}
\end{equation}
The angular correlation parameters
$A_v$, $A_t$ and $A_\Delta$ are given by:

\begin{equation}
A_v=1+\frac{1}{\overline{\Gamma}_0}
\Bigg[ 2\, {\rm Im}\Big[ \big( C_0(V)-L_0(V)\big)
                         \big( C_1(A)-L_1(A)\big)^{*} \Big]
-|M_1(V)-E_1(A)|^2\Bigg]
\ , \label{eq:av} 
\end{equation}
\begin{eqnarray}
A_t&=&{4\over 3}{1\over \overline{\Gamma}_0 }
\Bigg[ {\rm Im}\Big[ \big( C_0(V)-L_0(V)\big)
                     \big( C_1(A)-L_1(A)\big)^{*} \Big] \nonumber \\
&&-{1 \over \sqrt{2}} {\rm Im}\Big[ \big( C_0(V)-L_0(V)\big)
                                  \big( M_1(V)-E_1(A)\big)^{*} \Big] \nonumber \\
&&+\,{1 \over \sqrt{2}} {\rm Re}\Big[ \big( C_1(A)-L_1(A)\big)
                                      \big( M_1(V)-E_1(A)\big)^{*} \Big]
-{1\over 2}|M_1(V)-E_1(A)|^2\Bigg] \ , \label{eq:at} 
\end{eqnarray}
\begin{eqnarray}
A_\Delta&=&{2\over 3}{1\over \overline{\Gamma}_0 }
\Bigg[ \sqrt{2} \, {\rm Im}\Big[ \big( C_0(V)-L_0(V)\big)
                                 \big( M_1(V)-E_1(A)\big)^{*} \Big] \nonumber \\
&&-\sqrt{2}\, {\rm Re}\Big[ \big( C_1(A)-L_1(A)\big)
                            \big( M_1(V)-E_1(A)\big)^{*} \Big] \nonumber \\
&&+\, {\rm Im}\Big[ \big( C_0(V)-L_0(V)\big)
                    \big( C_1(A)-L_1(A)\big)^{*} \Big]
-{1\over 2}|M_1(V)-E_1(A)|^2\Bigg] \ . \label{eq:ad}
\end{eqnarray}
Finally, the coefficients $P_v$, $P_t$ and $P_\Delta$ are linear combinations
of the probabilities $P(f,\,f_z)$, and are defined
as~\cite{Hwa78,Con92}

\begin{eqnarray}
P_v      &=&P(1,1)-P(1,-1) \ , \nonumber \\
P_t      &=&P(1,1)+P(1,-1)-2\,P(1,0) \ , \nonumber \\
P_\Delta &=&P(1,1)+P(1,-1)+P(1,0)-3\,P(0,0)
\,=\,1-4\,P(0,0) \label{eq:pol} \ .
\end{eqnarray}
Therefore, $P_v$ and $P_t$ are proportional to the vector and tensor
polarizations of the $f$=1 state, respectively, while $P_{\Delta}$
indicates the deviation of the $f$=0 population density from its statistical
factor 1/4.  Because of the small energy splitting between the $f$=0 and $f$=1
hyperfine states (1.5 eV) compared to the $\mu^-\,^3$He binding energy,
and hence small deviation of $P(f,\,f_z)$ from its statistical value,
direct measurements of the angular correlation parameters are 
rather difficult~\cite{Mea01,Sou98,Con92}.
\section{The Weak Charge and Current Operators}
\label{sec:wcc}

An exhaustive description of the model for the nuclear weak
current has been recently given in Ref.~\cite{Mar01}.  Here
only its main features are summarized, and the new 
pseudo-scalar contributions are discussed. 

The nuclear weak current consists of vector and axial-vector parts, with
corresponding one- and two-body components.  The weak vector current
is constructed from the isovector part of the electromagnetic current,
in accordance with the conserved-vector-current (CVC) hypothesis.
One important difference between the present calculations and those
reported in Ref.~\cite{Mar01} is that the leptonic four-momentum transfer
is not negligible, but in fact close to the muon rest-mass.  Consequently,
electromagnetic form factors need to be included in the
expressions listed in Ref.~\cite{Mar01}.  The parameterization
used for these reproduces available $e$$N$ elastic scattering data.
Furthermore, in the present work the Darwin-Foldy relativistic correction to the
vector charge operator is also included. 

The one-body terms in the axial charge and current
operators have the standard expressions~\cite{Mar01} obtained
from the non-relativistic reduction of the covariant
single-nucleon current, and include terms proportional
to $1/m^2$, $m$ being the nucleon mass.  The induced 
pseudo-scalar contributions are retained both in the axial 
current and charge operators.  In particular, the pseudo-scalar 
axial charge operator is taken as

\begin{equation}
\rho_{i,PS}^{(1)}({\bf q};A)= -{g_{PS}\over{2\,m\,m_\mu}}\,(m_\mu-E_\nu)\, 
({\bbox \sigma}_i\cdot{\bf q})\,\tau_{i,-} \ ,
\label{eq:rhops}
\end{equation}
in the notation of Ref.~\cite{Mar01}.

Again, because the leptonic momentum transfer involved in muon
capture is not negligible, axial and induced pseudo-scalar form factors
need to be included.  These are parameterized as 

\begin{eqnarray}
g_A(q_\sigma^2) &=& { g_A \over {(1+ q_\sigma^2/\Lambda_A^2)^2} } \ ,
\label{eq:ga} \\
g_{PS}(q_\sigma^2) &=& -{2\,m_\mu\,m \over m_\pi^2 + q_\sigma^2 }\,
g_{A}(q_\sigma^2) \ ,
\label{eq:gps}
\end{eqnarray}
where $q_\sigma^2$ is the four-momentum transfer.
The axial-vector coupling constant $g_A$
is taken to be~\cite{Ade98} 1.2654$\pm$0.0042, by averaging values obtained
from the beta asymmetry in the decay of polarized neutrons and
the half-lives of the neutron and super-allowed $0^+ \rightarrow 0^+$
transitions.  The value for the cutoff mass $\Lambda_A$
is found to be approximately 1 GeV/$c^2$ from an analysis of
pion electro-production data~\cite{Ama79} and measurements of the
reaction $p$($\nu_\mu$,$\mu^+$)$n$~\cite{Kit83}.
The $q^2_\sigma$-dependence of $g_{PS}$ is obtained in accordance with 
the partially-conserved-axial-current (PCAC) hypothesis, by assuming
pion-pole dominance and the Goldberger-Treiman relation~\cite{Hem95,Wal75,Wal95},
$m_\pi$ here indicates the pion mass.

Some of the two-body axial-current operators are derived from
$\pi$- and $\rho$-meson exchanges and the $\rho\pi$-transition
mechanism.  These mesonic operators, first obtained in a systematic way
in Ref.~\cite{Che71}, give rather small contributions~\cite{Mar01}.
The two-body weak axial-charge operator includes a pion-range term,
which follows from soft-pion theorem and current algebra
arguments~\cite{Kub78,Tow92}, and short-range terms,
associated with scalar- and vector-meson exchanges. 
The latter are obtained consistently with the two-nucleon
interaction model, following a procedure~\cite{Kir92} similar
to that used to derive the corresponding weak vector-current
operators~\cite{Mar01}.  The two-body axial charge operator 
due to $N\Delta$-transition is also included~\cite{Mar01},
but its contribution is found to be very small.

The dominant two-body axial current operator is that due to
$\Delta$-isobar excitation~\cite{Sch98,Mar01}.  We briefly review here
its main features.  The $N\Delta$-transition axial current is written as
(notation as in Ref.~\cite{Mar01})

\begin{equation}
J_{i}^{(1)}({\bf q}; N\rightarrow\Delta, A) = 
         -\Bigg[g_A^*(q_\sigma^2) {\bf S}_i +
         \frac{g_{PS}^*(q_\sigma^2)}{2\,m\,m_\mu}{\bf q}
         ({\bf S}_i\cdot{\bf q})\Bigg]
         {\rm e}^{{\rm i}{\bf q}\cdot{\bf r}_i} T_{i,\pm} \ ,
\label{eq:j1ND}
\end{equation}
where ${\bf S}_i$ and ${\bf T}_i$ are spin- and isospin-transition 
operators, which convert a nucleon into a $\Delta$.  The induced
pseudo-scalar contribution, ignored in Ref.~\cite{Mar01}, has been 
obtained from a non-relativistic reduction of the covariant $N\Delta$-transition 
axial current~\cite{Hem95}. 

The axial and pseudo-scalar form factors $g_A^*$ and $g_{PS}^*$ 
are parameterized as 

\begin{eqnarray}
g_A^*(q_\sigma^2)&=& R_A\, g_A(q_\sigma^2) \ , \nonumber \\
g_{PS}^*(q_\sigma^2)&=&-{2\,m_\mu\,m \over m_\pi^2 + q_\sigma^2 }\,
g_{A}^*(q_\sigma^2) \ , 
\label{eq:gpsd}
\end{eqnarray}
with $g_A(q_\sigma^2)$ given in Eq.~(\ref{eq:ga}).
The parameter $R_A$ is adjusted to reproduce the experimental 
value of the Gamow-Teller matrix element in tritium $\beta$ 
decay, GT$^{\rm EXP}$ = 0.957 $\pm$ 0.003~\cite{Sch98}, 
while the $q_\sigma^2$-dependence of $g_{PS}^*$ is again 
obtained by assuming pion-pole dominance and PCAC~\cite{Hem95,Wal75,Wal95}. 
The values for $R_A$ determined in the present study are listed in
Table~\ref{tb:ga} for the four different combinations of interaction models.
The experimental error on GT$^{\rm EXP}$ is responsible for the
8--9 \% uncertainty in $R_A$.

Before concluding this section, a couple of remarks are in order.
First, it is important to note that the value of $R_A$ depends
on how the $\Delta$-isobar degrees of freedom are treated.
In the present work, the two-body $\Delta$-excitation operator is
derived in the static $\Delta$ approximation, using first-order
perturbation theory (see Ref.~\cite{Mar01}).  This approach
is considerably simpler than that adopted in Ref.~\cite{Mar01},
where the $\Delta$ degrees of freedom were treated non-perturbatively,
by retaining them explicitly in the nuclear wave functions~\cite{Sch92}.
The results for $R_A$ obtained within the perturbative (PT) and
non-perturbative (TCO) schemes differ by more than a factor of 2--see
Table VI of Ref.~\cite{Mar01}: $R_A$(PT)=1.22 and 
$R_A$(TCO)=2.87\footnote{Note that the value for $R_A$(PT) reported in 
Ref.~\cite{Mar01} is slightly different from that given 
here in Table~\ref{tb:ga}, since that value was obtained
from a random walk consisting of 100,000 configurations,
while the number of configurations sampled in
the present work is 150,000.}.
However, the results for the observables
calculated consistently within the two different schemes are typically
within 1 \% of each other.

Second, because of the procedure adopted to determine $R_A$,
the coupling constant $g_A^*$=$R_A\, g_A$ cannot be naively
interpreted as the $N$$\Delta$ axial coupling constant.  The
excitation of additional resonances and their associated
contributions will contaminate the value of $g_A^*$.  Indeed,
the PCAC arguments used above imply $g_A^*/g_A$=$f_{\pi N\Delta}/f_{\pi NN}$,
where $f_{\pi NN}$ and $f_{\pi N\Delta}$ are the 
$\pi$$N$$N$ and $\pi$$N$$\Delta$
coupling constants, and therefore one would obtain on the basis of 
Table~\ref{tb:ga}
that $(f_{\pi N\Delta}/f_{\pi NN})^2$ is in the range 1.08--1.56,
smaller than the value inferred from the $\Delta$
width, 4.67, and even smaller than the quark-model prediction, 2.88.

\section{Results}
\label{sec:res}

In this section results for the $^3$He($\mu^-$, $\nu_\mu$)$^3$H
capture process are reported.  The trinucleon
wave functions have been obtained from a realistic
Hamiltonian consisting of the Argonne $v_{18}$ (AV18)~\cite{Wir95}
two-nucleon and Urbana IX (UIX)~\cite{Pud95}
three-nucleon interactions.  To compare with
earlier predictions~\cite{Con92,Con95} for the
same process, and to have some estimate of the
model dependence, the older Argonne $v_{14}$ (AV14)~\cite{Wir84}
two-nucleon and Tucson-Melbourne (TM)~\cite{Coo79}
three-nucleon interactions have also been used.  Note
that both the UIX and TM interactions have been adjusted
to reproduce the triton binding energy.
Finally, to investigate the effect of the
three-nucleon interaction, predictions for muon-capture
observables have been made by including only two-nucleon
interactions (AV14 or AV18) in the Hamiltonian models.

The three-body bound-state problem has been solved with the
correlated-hyperspherical-harmonics (CHH) method, as described
in Refs.~\cite{Kie93,Kie94}.  
It consists essentially in expanding the wave function on the CHH basis,
and in determining variationally the expansion coefficients by
applying the Rayleigh-Ritz variational principle.

The $^3$H and $^3$He binding energies are listed in Table~\ref{tb:be}
for the different model Hamiltonians employed in the present
work.  They are obtained including only the isospin 1/2 
components of the wave functions.  
These results, which are very accurate 
(the uncertainty is of the order of one keV), 
are in excellent agreement with the values calculated 
using other techniques (for a review, see Ref.~\cite{Car98}).

Results for the capture rate $\Gamma_0$ and
angular correlation parameters $A_v$, $A_t$, and
$A_{\Delta}$, defined in Eqs.~(\ref{eq:g0})--(\ref{eq:ad}),
are presented in Table~\ref{tb:gapm}.  The uncertainty
(in parethesis) in the predicted values is due to
the uncertainty in the determination of  
the $N$$\Delta$ transition coupling constant $g_A^*$
(see Sec.~\ref{sec:wcc} and Table~\ref{tb:ga}).
The latter reflects the experimental error in
the Gamow-Teller matrix element of tritium
$\beta$-decay.

Inspection of Table~\ref{tb:gapm} shows that the
theoretical determination of the total capture rate $\Gamma_0$,
when the AV18/UIX and AV14/TM Hamiltonian models are used, 
is within 1 \% of the recent experimental result~\cite{Ack98}, $1496\pm 4$ sec$^{-1}$.
When the theoretical and experimental uncertainties are 
taken into consideration, the agreement between theory and
experiment is excellent.  Furthermore,
the model dependence in the calculated observables
is very weak: the AV18/UIX and AV14/TM results
differ by less than 0.5 \%.  The agreement between
theory and experiment and the weak model dependence
mentioned above reflect, to a large extent, the fact that
both the AV18/UIX and AV14/TM Hamiltonian models
reproduce: i) the experimental binding energies as well
as the charge and magnetic radii~\cite{Mar98} of the
trinucleons; ii) the Gamow-Teller matrix element in
tritium $\beta$-decay.  In this respect, it is interesting to note
that the capture rates predicted by the AV18 and
AV14 Hamiltonian models are about 4 \% smaller
than the experimental value, presumably
because of the under-prediction of the binding
energies and consequent over-prediction of the
radii. This makes the relevant nuclear
form factors, entering into the expression for
the rate $\Gamma_0$, smaller at the momentum transfer
of interest, $q \simeq 103$ MeV/$c$, than they
would be otherwise. To study how the rate $\Gamma_0$ scales with the 
triton binding energy, we have repeated the calculation using 
CHH wave functions obtained with a modified AV14/TM Hamiltonian 
model, which gives for the $^3$H and $^3$He binding energies 
9.042 and 8.349, respectively. The result for the rate $\Gamma_0$ 
is 1509 $\pm$ 7 sec$^{-1}$, while the angular correlation parameters 
are very close to the AV14/TM values listed in Table~\ref{tb:gapm}. 
Therefore, the rate $\Gamma_0$ scales 
approximately linearly with the trinucleon binding 
energy.  The values for the angular correlation
parameter $A_v$ listed in Table~\ref{tb:gapm}
can be compared with the experimental result
of Ref.~\cite{Sou98}, 0.63 $\pm$ 0.09 (stat.)$^{+0.11}_{-0.14}$ (syst.).
Theory and experiment are in agreement, for
any of the Hamiltonian models considered here.
However, the experimental uncertainity is much
larger than the theoretical one.

The contributions of the different components of
the weak current and charge operators to the
observables and to the RME's of the contributing
multipoles are reported for the AV18/UIX model
in Tables~\ref{tb:gac}, and~\ref{tb:cmc}--\ref{tb:mecc}, 
respectively.
The coupling constant $g_A^*$ has been set equal to
the central value of $1.17$ $g_A$ (see Table~\ref{tb:ga}).
The notation in Tables~\ref{tb:gac},~\ref{tb:cmc} and~\ref{tb:mecc} 
is as follows.  The column labeled \lq\lq One-body no PS\rq\rq
lists the contributions associated with the one-body
terms of the vector and axial charge and current
operators, including relativistic corrections
proportional to $1/m^2$.  However, the induced pseudo-scalar 
contributions are not considered in both the 
axial current and charge operators. 
Therefore, the \lq\lq One-body no PS\rq\rq contribution is 
associated with the operators given in Eqs.~(4.5)--(4.7),~(4.8),~(4.10),
and~(4.11)--(4.13) of Ref.~\cite{Mar01}, suitably modified
by the inclusion of nucleon form factors, as
explained in Sec.~\ref{sec:wcc}.
The column labeled \lq\lq One-body\rq\rq lists
the contribution obtained when, in addition, the induced 
pseudo-scalar axial charge and current operators, Eq.~(\ref{eq:rhops}) 
and last term of Eq.~(4.13) of Ref.~\cite{Mar01}, respectively,
are also included.

The column labeled \lq\lq Mesonic\rq\rq lists the results 
obtained by including, in addition, the 
contributions from two-body
vector and axial charge and current
operators, associated with pion-
and vector-meson-exchanges, i.e. 
the $\pi$V and $\rho$V for the vector current and charge operators, 
the $\pi$A, $\rho$A and $\rho\pi$A 
for the axial current operator, and 
the $\pi$A, sA and vA for the axial charge operator.
We have used the notation of Ref.~\cite{Mar01}, where 
these terms are listed respectively in 
Eqs.~(4.16)--(4.17),~(4.30)--(4.31),~(4.32)--(4.34)  
and~(4.35)--(4.37).
All these operators have been again
modified by the inclusion of form factors. 

The column labeled \lq\lq $\Delta$ no PS\rq\rq lists
the contributions arising from $\Delta$
excitation, but does not include those due to the induced 
pseudo-scalar $\Delta$ current of Eq.~(\ref{eq:j1ND}). 
The latter are retained in the column labeled
\lq\lq Full\rq\rq.  The associated operators
are obtained, as mentioned earlier
in Sec.~\ref{sec:wcc},  using perturbation
theory and the static $\Delta$ approximation
as in Eqs.~(4.44),~(4.48),~(4.50) and~(4.52) of Ref.~\cite{Mar01}.

Note that in Tables~\ref{tb:cmc} and~\ref{tb:mecc} 
the values for the RME $L_0(V)$ have not been listed, since the charge
and longitudinal multipole operators of the weak vector current,
denoted respectively as $C_{ll_z}(q;V)$ and $L_{ll_z}(q;V)$,
are related via CVC as~\cite{Mar01}

\begin{equation}
L_{ll_z}(q;V) = -\frac{1}{q}
\, \left[ H \, , \, C_{ll_z}(q;V) \right] \ .
\label{eq:lw1}
\end{equation}
In turn, this implies the following proportionality
between the corresponding RME's $C_0(V)$ and $L_0(V)$,
$L_0(V)=(m_\tau - m_t -q^2/2 m_t)\, C_0(V)/q$, or
$L_0(V)\simeq -0.024\, C_0(V)$ for $q \simeq 103$ MeV/$c$.
Finally, in Table~\ref{tb:cmc} the induced pseudo-scalar 
axial contributions are present only in $C_1(A)$ and $L_1(A)$, 
but not in $E_1(A)$, since the pseudo-scalar current is longitudinal.

The importance of the induced pseudo-scalar contribution 
can be understood by inspection of Table~\ref{tb:gac}. 
The nucleon induced pseudo-scalar term in the axial current and 
charge operators reduce the value of $\Gamma_0$ by about 16 \%, 
while the changes in the polarization observables are even larger. 
Far less important is the 
contribution from the pseudo-scalar $\Delta$-current, 
which reduces the value of $\Gamma_0$ by 
less than 1 \%. The changes in the polarization observables are also 
small, a few \%.

Among the observables, $\Gamma_0$ and $A_\Delta$
are the most sensitive to two-body contributions
in the weak current.  These are in fact crucial
for reproducing the experimental capture rate,
see Table~\ref{tb:gac}.  Inspection of Table~\ref{tb:cmc}
shows that two-body contributions are significant
in the RME's $M_1(V)$, $L_1(A)$, and $E_1(A)$, but
negligible in $C_0(V)$.  The $C_0(V)$ and $M_1(V)$
RME's are related by CVC to the corresponding RME's
of the isovector part of the electromagnetic current, since

\begin{equation}
{\bf j}_{-}({\bf q};V)=
 \Big[\, T_{-} \, , \, {\bf j}_{\rm iv}({\bf q};\gamma)\, \Big] \ ,
\end{equation}
where ${\bf j}_{-}({\bf q};V)$ is charge-lowering weak
vector current, ${\bf j}_{\rm iv}({\bf q};\gamma)$ is
the isovector part of the electromagnetic current, and
$T_{-}$ is the (total) isospin-lowering operator.  A similar
relation holds between the electromagnetic charge operator
and its weak vector counterpart.  Thus, if $^3$He and $^3$H
were truly members of an isospin doublet, then the $C_0(V)$
and $M_1(V)$ RME's would just be proportional to the isovector
combination of the trinucleon charge and magnetic form factors.
Of course, electromagnetic terms and isospin-symmetry-breaking
strong-interaction components in the nuclear potentials
spoil this property.  For example, the AV18/UIX model predicts
for the isovector RME's $C_{0,{\rm iv}}(\gamma)$ and
$M_{1,{\rm iv}}(\gamma)$ at $q \simeq 103$ MeV/$c$ 
the values 0.3250 and --0.1385 
(0.3254 and --0.1113 in impulse approximation), respectively.

The $C_1(A)$ RME is about two orders
of magnitude smaller than the leading RME's, as expected
on the basis of the following naive argument.  The
one-body axial charge density operator can be written
approximately as (the notation is that of Ref.~\cite{Mar01})
 
\begin{equation}
\rho^{(1)}_i({\bf x};A)=-\frac{g_A}{2\, m}\tau_{i,-}{\bbox \sigma}_i
\cdot \left[\, {\bf p}_i \, , \, \delta({\bf x}-{\bf r}_i) \, \right]_+
\simeq {\rm i} \frac{g_A}{2\, m} \tau_{i,-} {\bbox \sigma}_i \cdot \nabla_i
\delta({\bf x}-{\bf r}_i) \ ,
\label{eq:rden}
\end{equation}
where the term proportional to ${\bf p}_i$
has been neglected, and 
the identity $[ A \, , \, B]_+= [ A \, , \, B]_{-} + 2\, B\, A$ has
been used. Here $[ A \, , \, B]_\pm$ denote the anticommutator (+)
and commutator (--), respectively.  
We have also neglected the induced pseudo-scalar contribution. 
The one-body axial current density (its leading term) is 

\begin{equation}
{\bf j}^{(1)}_{i,{\rm NR}}({\bf x};A)=-g_A\, \tau_{i,-} \, {\bbox \sigma}_i\,
\delta({\bf x}-{\bf r}_i) \ ,
\label{eq:jden}
\end{equation}
and insertion of the approximation~(\ref{eq:rden}) and Eq.~(\ref{eq:jden})
into the expressions for the charge and longitudinal multipole operators
leads to the following relation between the associated RME's:
$C_1(A) \simeq -(q/2m)\, L_1(A)$, which, for $q \simeq 103$ MeV/$c$,
gives $C_1(A) \simeq -0.055 \, L_1(A)$, i.e.\ the correct sign
and order of magnitude obtained in the calculation. 

Lastly, from inspection of Table~\ref{tb:mecc}, 
it is interesting to note that
the contribution $\pi$A from the pion-exchange axial charge 
operator, which would be expected to be dominant among the 
two-body contributions to $C_1(A)$, is also negligible.
In fact, the operator
structure of the corresponding $C_1(A)$ multipole is such that it cannot
connect the dominant S-wave components in the $^3$He and $^3$H
wave functions, and the associated matrix element is therefore highly
suppressed.  Furthermore, the $\pi$A, $\rho$A and $\rho\pi$A contributions 
to $L_1(A)$ and $E_1(A)$ are about two orders of magnitude 
smaller than the leading one-body term (see Table~\ref{tb:gac}), 
and the relative signs of these contributions are such that 
they essentially cancel out in the total sum.  This feature of 
the mesonic contributions to the axial current was already found 
in other low-energy weak processes~\cite{Sch98,Mar00,Mar01}.

In order to compare with the results of Ref.~\cite{Con92},
the capture rate and angular correlation parameters
have been calculated with the CHH wave functions
corresponding to the AV14/TM Hamiltonian, and with
a model for the nuclear weak current including only
one-body terms.  The values for the coupling constants
and form factors entering the expressions for the
charge and current operators have been taken from
Ref.~\cite{Con92}.  The comparison between the present and
earlier predictions is shown in Table~\ref{tb:gacmp}:
there is satisfactory agreement between the two calculations.
The remaining 1--3 \% differences can presumably be 
explained as follows: i) the nuclear wave
functions have been obtained with an AV14/TM 
Hamiltonian model with slightly different cutoff 
parameters~\cite{Kampc}; ii) the weak one-body operators
in Ref.~\cite{Con92} include
some of the next-to-next-to-leading orders
in the non-relativistic expansion of the covariant
single-nucleon current, proportional to $1/m^3$, these are ignored
in the present calculation; iii) the numerical evaluation
of the required matrix elements is performed with different
techniques.  Here, Monte Carlo methods based on the
Metropolis {\it et al.} algorithm~\cite{Met53}
have been used.  Typically, the statistical error on the
calculated capture rate is less than 0.05 \%.

The results listed in Table~\ref{tb:gapm}, column labeled
\lq\lq AV14/TM\rq\rq, are also in good agreement
with those of Table~IX of Ref.~\cite{Con95}, although
the treatment of the short-range behavior of the
two-body terms in the weak current as well as
the values for the vector and axial form factors, coupling
constants, etc.~in Ref.~\cite{Con95} are slightly different
from those adopted in the present work.
It is important to emphasize, though, that the present model
for the weak current reproduces well the available
experimental data: i) the isovector component of the electromagnetic
current, which by CVC is related to the weak vector current, leads
to predictions for the isovector combination of the charge and
magnetic form factors of $^3$He and $^3$H in excellent agreement
with the measured values~\cite{Mar98} up to momentum transfer of
$\simeq 3$ fm$^{-1}$; ii) the two-body axial current operators
are constrained to reproduce the Gamow-Teller matrix element
in tritium $\beta$-decay.

To test the sensitivity of all the muon capture
observables to the induced pseudo-scalar form factors 
$g_{PS}$ and $g_{PS}^*$, Eqs.~(\ref{eq:gps}) and~(\ref{eq:gpsd}),
we have repeated the calculation 
using AV18/UIX CHH wave functions and several different 
values of $g_{PS}$ and $g_{PS}^*$ in terms of their 
PCAC predictions $g_{PS}^{\rm PCAC}$ and $g_{PS}^{\rm *\, PCAC}$.
We have assumed

\begin{equation}
R_{PS}\equiv \frac{g_{PS}}{g_{PS}^{\rm PCAC}}=
\frac{g_{PS}^*}{g_{PS}^{\rm *\, PCAC}} \ .
\label{eq:rps}
\end{equation}

\begin{figure}[bth]
\let\picnaturalsize=N
\def\picsize{5in}
\def\picfilenamea{mucap.gps.eps}
\ifx\nopictures Y\else{\ifx\epsfloaded Y\else\input epsf \fi
\let\epsfloaded=Y
\centerline{
\ifx\picnaturalsize N\epsfxsize \picsize\fi \epsfbox{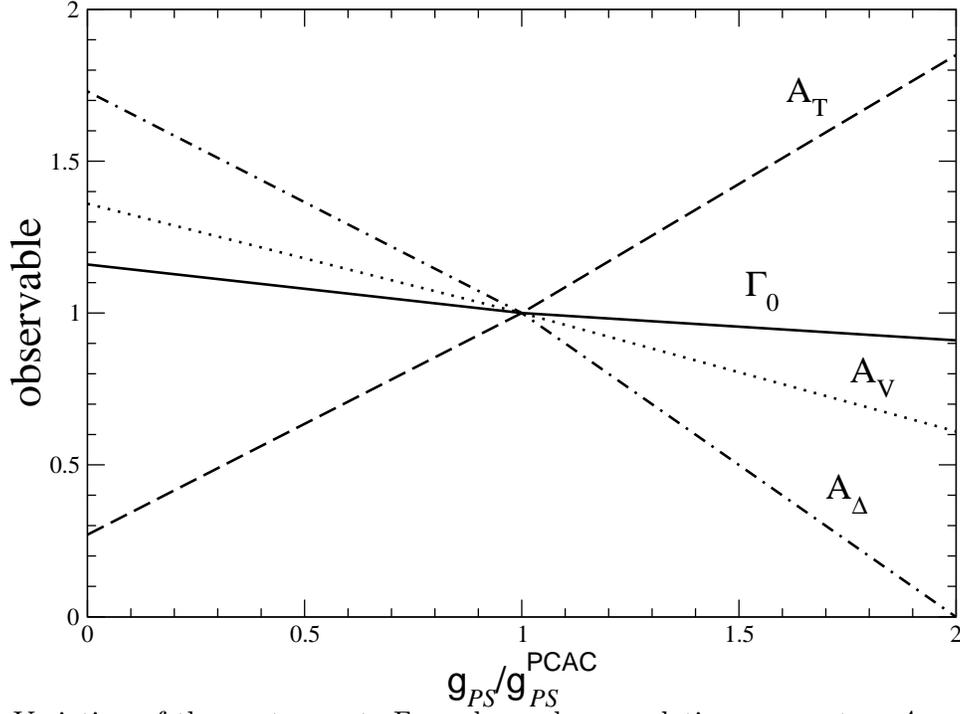}
 }}\fi
\caption{Variation of the capture rate $\Gamma_0$ and
angular correlation parameters $A_v$, $A_t$, and $A_\Delta$
with the induced pseudo-scalar coupling $g_{PS}$. The AV18/UIX
CHH wave functions are used.  For each observable, the ratio 
between the result obtained with the given value of $g_{PS}$
and the PCAC prediction, listed in Table~\protect\ref{tb:gapm},
is plotted {\it versus} the ratio
$g_{PS}/g_{PS}^{\rm PCAC}$ (=$g^*_{PS}/g_{PS}^{* {\rm PCAC}}$, see text).}
\label{fig:gps}
\end{figure}

The variation of each observable in terms of $R_{PS}$
is displayed in Fig.~\ref{fig:gps}.
The angular correlation parameters, in particular $A_t$ and
$A_\Delta$, are more sensitive to changes in $g_{PS}$ 
and $g_{PS}^*$ than
the total capture rate, as first pointed out in Ref.~\cite{Hwa78}.
A precise measurement of these polarization observables
could therefore be useful to ascertain the extent to which
the induced pseudo-scalar form factors deviate from their PCAC values.

Finally, by enforcing perfect agreement between the experimental and 
theoretical values, taken with their uncertainties, for the total
capture rate $\Gamma_0$, it is possible to obtain an estimate for
the range of values allowed for $R_{PS}$.
The procedure adopted is the following: i) we have considered the 
AV18/UIX minimum and maximum value for $\Gamma_0$ 
(see Table~\ref{tb:gapm}), obtained with $R_{PS}$=1 and $R_A$=1.08 
and 1.26, respectively (see Table~\ref{tb:ga}). ii) For these two 
values of $R_A$, we have tuned $R_{PS}$ to find $\Gamma_0$ 
within the experimental range. 
Our result for $R_{PS}$ is then  
\begin{equation}
R_{PS}=0.94 \pm 0.06 \ .
\label{eq:gpsth}
\end{equation} 
This 6 \% uncertainty is smaller than that found
in previous studies~\cite{Con92,Con95,Gov00}.  This
substantial reduction in uncertainty can be traced back
to the procedure used to constrain the (model-dependent)
two-body axial currents described in Sec.~\ref{sec:wcc}.
In this respect, it is interesting to note that ignoring altogether
the mesonic axial contributions associated with the $\pi$-, $\rho$-
and $\rho\pi$-exchange operators, and again re-adjusting the $N$$\Delta$
axial coupling constant to reproduce the tritium Gamow-Teller
matrix element (in this case, $g_A^*=1.32(9) g_A$ is required)
lead to the following predictions for the muon capture rate
and angular correlation parameters: $\Gamma_0$=1479(7) sec$^{-1}$,
$A_v$=0.5346(8), $A_t$=--0.3666(14), and $A_\Delta$=--0.0988(13).
In this case, the extracted value for the ratio $R_{PS}$
is 0.91 $\pm$ 0.06, in excellent agreement with the value of 
Eq.~(\ref{eq:gpsth}), suggesting that $R_{PS}$ is not
too sensitive to these mesonic contributions.
\section{Summary and Conclusions}
\label{sec:setc}

Muon capture observables for the process $^3$He($\mu^-$, $\nu_\mu$)$^3$H
have been calculated with very accurate CHH
wave functions corresponding to realistic Hamiltonians, the
AV18/UIX and AV14/TM models, and with a nuclear weak current
consisting of vector and axial-vector parts with one- and
two-body terms.  The conserved-vector-current hypothesis
has been used to derive the weak vector charge and
current operators from the isovector electromagnetic  
counterparts, while the axial current has been constructed
to reproduce the measured Gamow-Teller matrix element of
$^3$H $\beta$-decay.  The axial current also includes the 
nucleon and $\Delta$ induced pseudo-scalar current operators.
It should be emphasized that the
model adopted for the electromagnetic current provides
an excellent description of the $^3$He and $^3$H charge
and magnetic form factors~\cite{Mar98} at low and medium
values of momentum transfers.

The predicted total capture rate is in agreement
with the experimental value, and has been found to have only
a weak model dependence: the AV18/UIX and AV14/TM results differ
by less than 0.5 \%.  The weak model dependence can be traced
back to the fact that both Hamiltonians reproduce the binding
energies, charge and magnetic radii of the trinucleons, and 
the Gamow-Teller matrix element in tritium $\beta$-decay.

It is important to note that, if the contributions
associated with two-body terms in the axial
current were to be neglected, the predicted capture
rate would be 1316 (1318) sec$^{-1}$ with AV18/UIX (AV14/TM), 
and so two-body mechanisms
are crucial for reproducing the experimental value.
The present work demonstrates that the
procedure adopted for constraining these two-body
contributions leads to a consistent description of
available experimental data on weak transitions in
the three-body systems.  It also corroborates the
robustness of our recent predictions for the cross
sections of the proton weak captures on $^1$H~\cite{Sch98}
and $^3$He~\cite{Mar00,Mar01}, which were obtained
with the same model for the nuclear weak current.

Finally, it would be interesting to study the
$^3$He($\mu^-$,$\nu_\mu$)$n$$d$ and $^3$He($\mu^-$,$\nu_\mu$)$n$$n$$p$
processes, both of which have been investigated
experimentally in Ref.~\cite{Kuh94} and
theoretically in Ref.~\cite{Ski99}.  
Since the CHH method is suitable 
to solve for the three-body bound and
scattering states~\cite{Kie98}, the study of
these two processes is also possible.
Work along these lines is vigorously being pursued.
\section*{Acknowledgments}
The authors wish to thank E.\ Truhl\`{\i}k
for useful discussions.  The work of R.S.\ is supported by 
the U.S. Department of Energy contract number DE-AC05-84ER40150
under which the Southeastern Universities Research Association (SURA)
operates the Thomas Jefferson National Accelerator Facility.  Some
of the calculations were made possible by grants of computing time from
the National Energy Research Supercomputer Center in Livermore.

%
%
%
\begin{table}
\caption{\label{tb:ga}
Values for $R_A=g_A^*/g_A$, where $g_A^*$ is 
the $N$$\Delta$ transition axial
coupling constant (see, however, Sec.~III for
a discussion of the proper interpretation
of $g_A^*$).  The results are 
obtained by reproducing the experimental value
of the Gamow-Teller (GT) matrix element in
tritium $\beta$-decay with CHH wave functions
corresponding to the AV18, AV14, AV18/UIX, and
AV14/TM Hamiltonian models.  The theoretical
uncertainties are due to the experimental error
with which the GT matrix element is known.
}
\begin{tabular}{cc}
Interaction Model & ${g_A^*/g_A}$  \\
\tableline
AV18     & 1.25$\pm$0.10 \\
AV14     & 1.11$\pm$0.09 \\
AV18/UIX & 1.17$\pm$0.09 \\
AV14/TM  & 1.04$\pm$0.09 \\
\end{tabular}
\end{table}
\begin{table}
\caption{\label{tb:be}
Binding energies in MeV of $^3$He and $^3$H calculated
with the CHH method using the AV18, AV14, AV18/UIX, 
and AV14/TM Hamiltonian models.  Also listed are
the experimental values. 
}
\begin{tabular}{ccc}
Interaction Model & $^3$He & $^3$H  \\
\tableline
AV18     & 6.917 & 7.617 \\
AV14     & 7.032 & 7.683 \\ 
AV18/UIX & 7.741 & 8.473 \\
AV14/TM  & 7.809 & 8.485 \\
\tableline
EXP      & 7.72  & 8.48  \\
\end{tabular}
\end{table}
\begin{table}
\caption{\label{tb:gapm} 
Capture rate $\Gamma_0$ in sec$^{-1}$, and angular
correlation parameters $A_v$, $A_t$, and $A_\Delta$,
as defined in Eqs.~(\ref{eq:g0})--(\ref{eq:ad}),
calculated using CHH wave functions corresponding
to the AV18, AV14, AV18/UIX, and AV14/TM Hamiltonian
models.  The theoretical uncertainties, shown in
parenthesis, reflect the uncertainty in the
determination of the $N$$\Delta$ transition
axial coupling constant $g_A^*$.
}
\begin{tabular}{ccccc}
Observable &    AV18     &   AV14     &  AV18/UIX    & AV14/TM      \\ 
\tableline
$\Gamma_0$ &  1441(7)    &  1444(7)    &  1484(8)    &  1486(8)   \\
$A_v$      &  0.5341(14) &  0.5339(14) &  0.5350(14) &  0.5336(14)\\
$A_t$      &--0.3642(9)  &--0.3643(9)  &--0.3650(9)  &--0.3659(9) \\ 
$A_\Delta$ &--0.1017(16) &--0.1018(16) &--0.1000(16) &--0.1005(17)\\ 
\end{tabular}
\end{table}
\begin{table}
\caption{\label{tb:gac} Cumulative contributions
to the capture rate $\Gamma_0$ (in sec$^{-1}$) 
and angular correlation parameters $A_v$, $A_t$,
and $A_\Delta$.  The CHH wave functions are obtained
using the AV18/UIX Hamiltonian model.  The column 
labeled \lq\lq One-body-no PS\rq\rq~lists the contributions
associated with the one-body vector and axial charge
and current operators, but no induced pseudo-scalar 
axial term is included. This is done in the column 
labeled \lq\lq One-body\rq\rq, while the column labeled 
\lq\lq Mesonic\rq\rq~lists the results obtained
by including, in addition, the contributions from
meson-exchange mechanisms.  Finally the column labeled
\lq\lq$\Delta$-no PS\rq\rq~lists the results obtained by including
also the $\Delta$-excitation contributions, 
with $g_A^*/g_A$ set to the central value of 1.17 
(see Table~\protect\ref{tb:ga}), but excluding the $\Delta$ 
pseudo-scalar term, which is included in the column 
labeled \lq\lq Full\rq\rq.}
\begin{tabular}{cccccc}
Observable & One-body no PS & One-body & Mesonic & $\Delta$ no PS &  Full   \\
\tableline
$\Gamma_0$ &  1530          &  1316    &  1384   &   1493         &  1484   \\
$A_v$      &  0.7735        &  0.5749  &  0.5511 &   0.5438       &  0.5350 \\
$A_t$      &--0.0840        &--0.3565  &--0.3679 & --0.3525       &--0.3650 \\ 
$A_\Delta$ &--0.1424        &--0.0686  &--0.0810 & --0.1038       &--0.1000 \\ 
\end{tabular}
\end{table}
\begin{table}[t]
\caption{\label{tb:cmc} Cumulative contributions to the
reduced matrix elements (RME's) $C_0(V)$, $C_1(A)$, $L_1(A)$,
$E_1(A)$, and $M_1(V)$.  The CHH wave functions are obtained
using the AV18/UIX Hamiltonian model.  Note that $C_0(V)$ is
purely real, while the other RME's are purely imaginary.
Notations as in Table~\protect\ref{tb:gac}.
}
\begin{tabular}{cccccc}
RME      & One-body no PS & One-body & Mesonic & $\Delta$ no PS &  Full   \\
\tableline 
$C_0(V)$ &   0.3280                  & &    0.3277                  & & \\
$C_1(A)$ & --0.7532$\times\,10^{-2}$ &--0.4076$\times\,10^{-2}$ 
         & --0.4135$\times\,10^{-2}$ &--0.4397$\times\,10^{-2}$  & \\ 
$L_1(A)$ &   0.4058  & 0.2590                  
         &   0.2618  & 0.2804 & 0.2737 \\
$E_1(A)$ &   0.5519                  & &  0.5563                  
         &   0.5813                  & \\ 
$M_1(V)$ & --0.1128                  & & --0.1314                  
         & --0.1355  &  \\
\end{tabular}
\end{table}
\begin{table}[t]
\caption{\label{tb:mecc} Individual mesonic contributions to the
reduced matrix elements (RME's) $C_0(V)$, $C_1(A)$, $L_1(A)$,
$E_1(A)$, and $M_1(V)$.  The CHH wave functions are obtained
using the AV18/UIX Hamiltonian model.  Note that $C_0(V)$ is
purely real, while the other RME's are purely imaginary.
Notations as explained in the text.
}
\begin{tabular}{cccccc}
RME      & $\pi$(V/A)   & $\rho$(V/A)  & $\rho\pi$A & sA & vA  \\
\tableline 
$C_0(V)$ & 
--0.3285$\times\,10^{-3}$ & --0.6950$\times\,10^{-4}$ &  &  &   \\
$C_1(A)$ & 
--0.3253$\times\,10^{-5}$ &  &  & --0.2730$\times\,10^{-3}$ & 
0.2174$\times\,10^{-3}$ \\ 
$L_1(A)$ & 
0.2324$\times\,10^{-2}$ & --0.2894$\times\,10^{-2}$ & 
0.3409$\times\,10^{-2}$ &  &  \\   
$E_1(A)$ & 
0.2539$\times\,10^{-2}$ & --0.4208$\times\,10^{-2}$ & 
0.6056$\times\,10^{-2}$ &  &  \\   
$M_1(V)$ & 
--0.1597$\times\,10^{-1}$ & --0.2627$\times\,10^{-2}$ &  &  &  \\
\end{tabular}
\end{table}
\begin{table}
\caption{\label{tb:gacmp} Capture rate $\Gamma_0$ (in sec$^{-1}$) 
and angular correlation parameters $A_v$, $A_t$, and $A_\Delta$
obtained with AV14/TM CHH wave functions, and only
one-body operators (column labeled \lq\lq One-body\rq\rq) are
compared with the results 
of Table~3 of Ref~\protect\cite{Con92}.
}
\begin{tabular}{ccc}
Observable &        One-body     & Ref.~\protect\cite{Con92}     \\
\tableline
$\Gamma_0$ &          1287        &       1304                     \\
$A_v$      &         0.579        &      0.568                     \\
$A_t$      &       --0.351        &    --0.356                     \\ 
$A_\Delta$ &       --0.070        &    --0.076                     \\ 
\end{tabular}
\end{table}
%
%
%
%
%
\end{document}